\documentclass[twocolumn,showpacs,amsmath,amssymb]{revtex4}
\usepackage{dcolumn}
\usepackage{bm}
\usepackage{graphicx}
\usepackage{color}

\begin{document}

\title{Electron-electron interaction effects in quantum point contacts}
\author{Anders Mathias Lunde,$^1$ Alessandro De Martino,$^2$ Reinhold Egger,$^2$ and Karsten Flensberg$^1$}
\affiliation{
$^1$~Nano-Science Center, Niels Bohr Institute, University of Copenhagen,
DK-2100 Copenhagen, Denmark \\
$^2$~Institut f\"ur Theoretische Physik, Heinrich-Heine-Universit\"at,
D-40225  D\"usseldorf, Germany}
\date{\today}

\begin{abstract}
We consider interaction effects in quantum point contacts on the
first quantization plateau, taking into account all non
momentum-conserving processes. We compute low-temperature linear and
non-linear conductance, shot noise, and thermopower by perturbation
theory, and show that they are consistent with experimental
observations on the so-called "0.7 anomaly". The full
temperature-dependent conductance is obtained from self-consistent
second-order perturbation theory  and approaches $\approx e^2/h$  at
higher temperatures, but still smaller than the Fermi temperature.
\end{abstract}

\maketitle

Conductance quantization in a quantum point contact (QPC), first
observed in 1988 \cite{wharam}, constitutes a classic textbook
effect of mesoscopic physics.  On top of the integer conductance
plateaus $G=nG_0$ (where $G_0=2e^2/h$) observed as a function of
gate voltage $V_g$, many experiments have pointed to the existence
of a so-called "0.7 anomaly" in the conductance and other transport
quantities \cite{thomas,kristensen,cronenwett,appleyard,roche}. Most
prominently, the 0.7 anomaly implies a shoulder-like feature in the
conductance $G(V_g)$ around $G\approx 0.7\,G_0$ seen at elevated
temperature $T$ (or finite voltage $V$) near the first quantized
plateau \cite{thomas,kristensen,cronenwett}, accompanied by a shot
noise reduction \cite{roche}. Given the conceptual simplicity of a
QPC and the fact that the 0.7 anomaly has been observed in a variety
of material systems by different groups over more than a decade, it
is quite amazing that still no generally accepted microscopic theory
exists, apart from an overall consensus that one is dealing with
some spin-related many-body effect. Such a theory should be able to
explain all the experimental data in a unified and physically
consistent manner.

While phenomenological models \cite{bruus}, assuming the existence
of a density-dependent spin gap, can provide rather good fits to
experimental data, the presumed static spin polarization due to
interactions within the {\sl local}\ QPC region is not expected in
the presence of unpolarized {\sl bulk}\ reservoirs. Recently it was
also pointed out that spin symmetry-broken mean-field theory is
unable to recover the correct $T$ dependence of the conductance
\cite{richter}.  Other proposals assume the existence of a
quasi-bound state in the QPC region, leading to a Kondo-type
scenario as encountered in transport through interacting quantum
dots \cite{meir,cornagliabalseiro}. Such a quasi-bound state was
indeed found in spin density functional theory (SDFT) calculations
\cite{meir}, but other SDFT works did not reach such conclusions
\cite{berggren}. Further proposals involve phonon effects
\cite{seeligmatveev}. Several publications have suggested that
taking into account only electron-electron (e-e) interactions may
result in a reduced conductance at elevated temperatures, without
the need for additional assumptions of spin polarization or a
localized state
\cite{matveev,meidanoreg,schmeltzer,syljuaasen,sushkov}. However, a
physically consistent picture explaining the temperature, voltage,
and magnetic field dependence of the conductance, as well as
thermopower and noise experiments, is still lacking. In this paper,
we show that a careful consideration of {\sl non
momentum-conserving} e-e interaction processes in QPCs may allow for
a consistent theory of the 0.7 anomaly.

\begin{figure}
\includegraphics[width=0.45\textwidth]{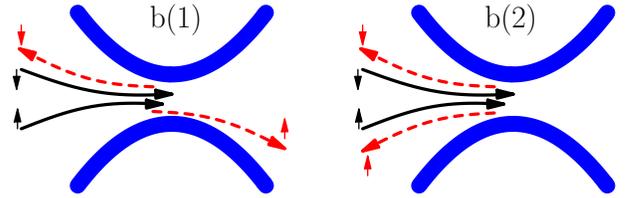}
\caption{ \label{fig1} (Color online) Illustration of the
two-electron non momentum-conserving scattering processes that give
rise to a correction to the transport properties at the beginning of
the first plateau. The full (black) lines represent incoming
electrons, while the dashed (red) lines are the outgoing electrons.
The thick (blue) lines define the edge of the QPC. Only scattering
between different spins is present to leading order in $T/T_F$ due
to the Pauli principle.}
\end{figure}

The lack of momentum conservation in e-e scattering processes is due
to an emerging lack of translational invariance relevant for the low density
regime $k_FL\sim 1$, where $k_F$ is the local Fermi momentum and $L$
a typical interaction length-scale of the QPC (see below). Therefore
this effect is dominant at the onset of the plateau and gradually
disappears for larger electron density in the QPC. We will focus on
the first conductance plateau, where the QPC has only one open
channel (1D mode) \cite{glazman88,adia}. Now e-e interactions give
the contribution
\begin{equation}\label{hi}
H_I = \frac12 \sum_{\sigma\sigma'} \int dx dx' \ W(x,x') \
\Psi_\sigma^\dagger(x) \Psi_{\sigma'}^\dagger(x') \Psi_{\sigma'}^{}
(x') \Psi_\sigma^{} (x)
\end{equation}
to the Hamiltonian, where $\Psi_\sigma(x)$ is the 1D electron field
operator for spin $\sigma=\uparrow,\downarrow$. The pair potential
$W(x,x')$ takes into account screening processes due to closed
channels and nearby gates, as well as semiclassical slowing down
\cite{sushkov}, and therefore depends not only on the relative
coordinate $x-x'$ but also on the center of mass $X=(x+x')/2$. In
fact, the range of the interaction $W(X)$ is determined by the above
length-scale $L$. The e-e interaction now enters the 1D description
in two different ways. There are (i) one-particle effects described
through a self-consistent potential, given by the real part of the
self energy (in the simplest case, this is the Hartree-Fock
potential). At $T=0$, the imaginary part of the self energy for our
Fermi liquid starting point is zero \cite{luttinger}, and the
plateau then occurs where the open channel suffers no
backscattering.  This part is thus already contained in the
potential forming the QPC. In addition, at finite $T$, we have (ii)
inelastic e-e scattering processes, which are the focus of our work
and give rise to a reduced conductance by changing the number of
left- and right-movers, and hence the current. This is illustrated
in Fig.~\ref{fig1}, where two important types of non
momentum-conserving scattering events are illustrated.  Process b(2)
describes the simultaneous backscattering of two electrons with
opposite spins and has been discussed on a perturbative level in
Ref.~\cite{meidanoreg}. The process b(1), where a single electron is
backscattered, has not been studied before. On top of these two,
there are e-e forward scattering and backscattering processes
(momentum-conserving in a long 1D wire).  While high-$T$ transport
properties are affected by all interactions, we show now that the
leading low-$T$ behavior is {\sl fully} determined by the two
processes in Fig.~\ref{fig1}.

Let us calculate the non-linear conductance, the thermopower, and
the shot noise to leading order in the interaction. Perturbation
theory gives the {\sl interaction correction to the current} (here,
$\hbar=1$ and $e>0$)
\begin{eqnarray}\label{currpert}
 \frac{I(V,T)}{G_0 V} &=&
1-   \left(A_{b(1)}+A_{b(2)}\right) (\pi T/T_F)^2
\\  \nonumber && - \left(A_{b(1)}/4+A_{b(2)}\right) (e V/\varepsilon_F)^2
+ {\cal O}(W^3),
\end{eqnarray}
with coefficients
 $A_{b(1,2)}= W_{b(1,2)}^2 k_F^4/(48\pi^2 \varepsilon_F^2)$
corresponding to the b(1) and b(2) processes, where
$\varepsilon_F=k_B T_F$ is the Fermi energy and $W_{b(1,2)} =\int
dxdx'\ W(x,x') e^{ik_F(x+x')}e^{ik_F(x \mp x')}$ . Already at this
point, we observe a correspondence to experimental observations,
namely the reduced conductance $I/V$ with increasing $T$ and/or $V$.
Furthermore, to leading order in $T/T_F$, where all scattering
happens at the Fermi level, only opposite spins interact due to the
Pauli principle: for equal spins, the exchange term tends to cancel
the direct term. As a consequence, we can also understand the
behavior at \textit{large magnetic fields}, where the $T=0$ plateau
occurs at $e^2/h$. In that case, to order $(T/T_F)^2$ no interaction
renormalization of the conductance arises \cite{Binfty}. This is
consistent with experiments, where no suppression is observed at the
half-plateaus.

Another experimental observable probing the enhanced phase space for
e-e scattering at higher $T$ is the {\sl thermopower}
$\mathcal{S}(T)$ \cite{thermodef}, for which perturbation theory
predicts
\begin{equation}\label{thermo}
\mathcal{S}(T) = \frac{k_B}{e} \frac{2\pi^4}{5}
\left(A_{b(1)}+A_{b(2)}\right)(T/T_F)^3 .
\end{equation}
Since the non-interacting thermopower is exponentially small
[$\propto \exp(-T_F/T)$] at the conductance plateau, the interaction
correction completely determines the low-temperature thermopower
\cite{lundeprl}. The enhanced thermopower (as compared to the
non-interacting one) is in qualitative agreement with
experiments at the anomalous plateau \cite{appleyard}.

Next we calculate consequences for another observable, namely
non-equilibrium noise. The zero-frequency {\sl shot noise} follows
from the (symmetrized) two-point correlation function of the current
operator.
Perturbation theory yields for the backscattering noise power
\begin{eqnarray}\label{noise}
S_B(V,T) &=& 2 e \ \Bigl[ 2  I_{bs(2)}(V,T) \coth(eV/k_B T) \\ \nonumber
& + & I_{bs(1)}(V,T)\coth(eV/2k_B T) \Bigr],
\end{eqnarray}
where $I_{bs(1,2)}$ are the current corrections due to $W_{b(1,2)}$
quoted in Eq.~(\ref{currpert}) (defined positive for $V>0$). This is
nothing but the famous Schottky shot noise relation, encoding the
charge of the backscattered particles. Equation (\ref{noise})
predicts an additional factor of two for the b(2) contribution,
because two electrons are backscattered in that event
\cite{meidanoreg}. Direct calculation then yields the full noise
power of the transmitted current as $S_T=S_B+4G_0k_BT-8k_BT
\partial_V I_{bs}$, where $I_{bs}=I_{bs(1)}+I_{bs(2)}$. Recent noise
measurements on the first quantized plateau were compared to the
corresponding single-particle picture \cite{roche}, and a reduced
noise power was observed on the conductance anomaly. For that
comparison, one subtracts the thermal noise and defines the excess
noise as $S_I=S_T-4G(V,T)k_BT$. For a non-interacting system,
$S_I^{SP}=2G_0R\{eV\coth(eV/2k_BT)-2k_BT\}$ to lowest order in the
reflection coefficient $R=I_{bs}/G_0 V$, see Ref.~\cite{roche}. Thus
the difference between the true excess noise and its single-particle
value is
\begin{equation}\label{diff}
   \frac{S_I- S_I^{SP}}{2G_0 eV (T/T_F)^2}=
   -2A_{b(1)}\frac{eV}{k_B T} + A_{b(2)}\, h(eV/k_BT),
\end{equation}
where $h(x)=-8x+(\pi^2+x^2)\tanh(x/2)$. This expression shows that
for $eV<$ 6.507 $k_BT$, regardless of $A_{b(1,2)}$, the measured
noise is always smaller than predicted by a single-particle
analysis. This situation corresponds to the experimental work of
Ref.~\cite{roche}, where $eV\lesssim 5 k_BT$.

It is clear from all these perturbative results that for low energies,
$V,T\to 0$, all interaction effects disappear. The perturbation
theory results presented above, however, obviously break down at
higher temperatures or voltages. From
Eq.~(\ref{currpert}), we find the temperature scale
for this crossover to a strong-interaction regime,
\begin{equation}\label{Tstar}
    k_BT^*\approx \frac{\varepsilon_F}{\sqrt{A_b}}\propto
    \frac{\varepsilon_F^2}{W_bk_F^2},
\end{equation}
where $A_b=A_{b(1)}+A_{b(2)}$ and
$W^2=W_{b(1)}^2+W_{b(2)}^2$. Contrary to the usual situation
encountered in mesoscopic physics, the nontrivial question to be
answered thus concerns the {\sl high-temperature} limit (but still
$T\ll T_F$). To make
progress in the relevant temperature regime
\begin{equation}\label{temprange}
    T^*\lesssim T\ll T_F,
\end{equation}
let us consider a simplified model pair potential entering Eq.~(\ref{hi}),
see Ref.~\cite{sushkov},
\begin{equation} \label{localpair}
W(x,x') =V_0 \delta(x) \delta(x'),
\end{equation}
which implies $W_{b(1)}=W_{b(2)}$. We express the interaction
strength in the dimensionless parameter $\lambda= mV_0/2\pi^{3/2}$.
Estimates for $\lambda$ in GaAs heterostructures gives
$\lambda\approx 1$, see Refs.~\cite{sushkov,richter}, which then
yields $k_BT^*/T_F\approx 0.1$ allowing for a study of the
temperature range \eqref{temprange}.

\begin{figure}
\includegraphics[width=0.45\textwidth]{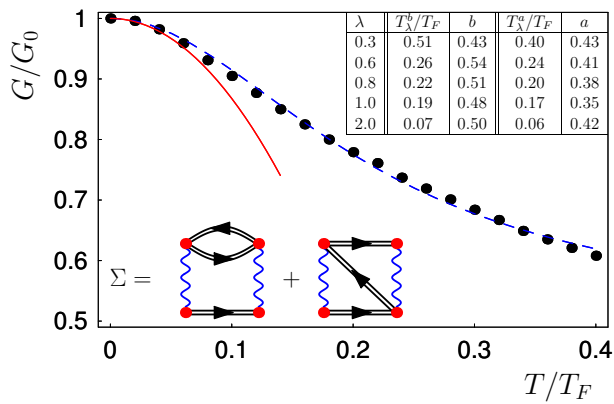}
\caption{ (Color online) \label{fig2} Temperature dependence of
 the linear conductance $G$ for $\lambda=0.8$. Dots denote
self-consistent numerical results, the solid curve gives the
perturbative estimate (\ref{currpert}), and the dotted curve is a
fit to Eq.~(\ref{boltzmann}). Inset: Fit parameters entering best
fits of Eqs.~(\ref{boltzmann}) and (\ref{activ}) to numerical data
for $T/T_F< 0.4$. Note that $a$ and $b$ are somewhat different. The
lower inset shows the two second order self-consistent energy
diagrams in Eq.~\eqref{selfen}. }
\end{figure}

In order to treat the local interaction (\ref{localpair}), we start
from the Dyson equation for the full Keldysh single-particle Green's
function (GF) $\mathbf{G}(x,x';\omega)$,
\begin{equation}\label{dyson}
\mathbf{G}(x,x'; \omega)= \mathbf{G}_0(x,x';\omega)+
\mathbf{G}_0(x,0;\omega) \mathbf{\Sigma}(\omega)
\mathbf{G}(0,x';\omega),
\end{equation}
which is a $2\times 2$ matrix in Keldysh space. The self energy due
to Eq.~(\ref{localpair}) acts only at $x=x'=0$. Note that
$\mathbf{G}_\uparrow=\mathbf{G}_\downarrow$, i.e., we can suppress
the spin index. For simplicity, we now assume a parabolic
dispersion, $\varepsilon_k=k^2/2m$, for the open channel. The charge
current operator is $I = \frac{ie}{2m} \sum_\sigma \left[
\Psi_\sigma^\dagger(x)
\partial_x \Psi^{}_\sigma(x) - (\partial_x \Psi^\dagger_\sigma(x))
\Psi_\sigma(x) \right]$,
and we evaluate $\langle I \rangle$ at $x=0$, where it can be
expressed in terms of the local GF $\mathbf{G}(\omega)\equiv
\mathbf{G}(0,0;\omega)$ and the self energy $\mathbf{\Sigma}(\omega)$.
In fact, some
algebra shows that only the local spectral function
$A(\omega)=-2 \ {\rm Im} G^r(\omega)$ enters the current formula for the
contact interaction (\ref{localpair}),
\begin{equation}\label{current}
I = \frac{2e}{h} \int_0^\infty d\omega
\left[f_R^0(\omega)-f_L^0(\omega)\right]
\frac{A(\omega)}{A_0(\omega)},
\end{equation}
where $f_{R/L}^0$ are Fermi functions in the right/left lead, and
$A_0(\omega)=2 \pi d(\omega)$ is the non-interacting spectral
function.  Here, $d(\omega)$ is the density of states $2\pi
d(\omega)= (2m/\omega)^{1/2}\theta(\omega)$. Remarkably, the
nonequilibrium current through the interacting QPC is thereby fully
expressed in terms of the local retarded GF only. So far, the given
relations are exact, but to make progress, one needs to approximate
the self energy. We take the full {\sl second-order self energy},
\begin{eqnarray}\label{selfen}
\nonumber\Sigma^r(\omega) &=& V_0^2 \int_0^\infty dt\ e^{i\omega t}
\left[
G^<(-t) G^>(t) G^>(t)\right. \\
 && \left.\quad- G^>(-t) G^<(t) G^<(t) )\right],
\end{eqnarray}
and make it self-consistent by using the interacting
(lesser/greater) GFs. The corresponding diagrams are shown in the
inset to Fig.~\ref{fig2}.  The approximation (\ref{selfen}) is the
simplest way to describe equilibration between left- and
right-moving electrons in an interacting QPC.
In what follows, we
confine ourselves to the linear conductance regime, where the
spectral function in Eq.~\eqref{current} can be calculated in
equilibrium by solving Eq.~\eqref{selfen}, and where we can replace
$f_R^0-f_L^0\to eV[-\partial_\omega f(\omega)]$, where $f(\omega)$
is the Fermi function.
 In linear response, the
lesser/greater GFs can be written in terms of the local spectral
function $A(\omega)$,
\begin{equation}\label{Glessgreat}
G^{</>}(t)= \pm i \int_0^\infty \frac{d\omega}{2\pi} \ e^{-i\omega
t} A(\omega) f(\pm \omega).
\end{equation}
This suggests a natural iterative way to self-consistently solve for
the conductance: Starting with the initial guess
$A(\omega)=A_0(\omega)$, one computes $\Sigma^r(\omega)$ from
Eq.~(\ref{selfen}), which in turn defines a new retarded GF and a
new guess for $A(\omega)$. This procedure is iterated until
convergence has been reached. For the parameters below, this
numerical scheme is convergent and can be implemented in an
efficient manner.

The numerical results, for $\lambda=0.8$ shown in  Fig.~\ref{fig2},
accurately reproduce the above perturbative results at low $T$, but
also allow to cover the interesting high-temperature limit. Our data
for different $\lambda$ fall to high accuracy on the simple function
\begin{equation}\label{boltzmann}
\frac{G(T)}{G_0} = b+ \frac{1-b}{1+(T/T_{\lambda}^b)^2} ,
\end{equation}
where $b$ sets the high-temperature saturation value. While this
functional dependence is somewhat similar to the phenomenological
Kondo-type function used in Ref.~\cite{cronenwett}, our numerical
data fit better to Eq.~(\ref{boltzmann}). It is also possible to
obtain equally good fits to the activated $T$ dependence reported in
Ref.~\cite{kristensen}, see also Ref.~\cite{bruus},
\begin{equation}\label{activ}
\frac{G(T)}{G_0} = 1 -(1-a) e^{-T_{\lambda}^a/T},
\end{equation}
where $a$ again denotes the high-$T$ limit. The values for
$T_\lambda^b$ and $T^a_\lambda$ extracted from best fits to our
numerical data are summarized in the inset in Fig.~\ref{fig2}.
Remarkably, both temperature scales are of the same order. Moreover,
they are lowered by increasing the interaction strength $\lambda$.
For high $T$, the conductance appears to approach the saturation
value $G\approx e^2/h$. Similar saturation value has also been
reported for \textit{long} wires \cite{matveev}, with the same $T=0$
conductance $G_0$. The new feature for QPCs comes from the
non-momentum conserving interactions, resulting in a distinct
low-to-intermediate temperature dependence $G(T)$. The perturbative
$T^2$ correction is not present in the long wire results
\cite{matveev}, but is seen experimentally \cite{cronenwett}.

As a final remark on the numerical solution of the self-consistent
approach, we mention that thermopower (data not shown) exhibits a
crossover from the ${\cal S}\propto T^3$ law at low $T$, see
Eq.~\eqref{thermo}, to a linear-in-$T$ behavior at elevated
temperatures.

It is also instructive to discuss our model in terms of an Anderson
model. For the model pair potential (\ref{localpair}), by spatial
discretization our Hamiltonian maps to a 1D tight-binding chain with
hopping matrix elements $t$ and on-site interaction $U$ acting at
one site ($x=0$) only \cite{footnew}. We thus arrive at an
Anderson-type impurity model, similar to the one used in
Ref.~\cite{meir} to describe interactions in a QPC in the Kondo
regime. However, we consider a rather different parameter regime,
where $U$ is of the same order as the hybridization $\Gamma$ and can
be parametrically larger than the bandwidth $D\sim |t|$. Employing
Eq.~(\ref{Tstar}), with $\varepsilon_F\approx D$, the interesting
temperature range \eqref{temprange} translates to $D^2/U \ll k_B
T\ll D$, where our claim is that $G$ approaches $\approx e^2/h$.
While the Kondo model requires the formation of a local moment, this
is not the case for the present approach. Instead, our
high-temperature limit may be described as an {\sl incoherent Fermi
liquid}, with {\sl full relaxation} between left- and right-movers.
In fact, one can establish that a simple Boltzmann-type approach has
a high-temperature solution where the out-going distribution
function is a mixture of the incoming left- and right-mover's
distributions. Such an Ansatz leads to the conductance formula
\eqref{boltzmann} with $b=1/2$, which is the reason for using that
as a fitting formula. Unfortunately, a Boltzmann approach is
conceptually difficult to justify due to an inherent normalization
problem\cite{sushkov}, i.e. one cannot define a  proper local
distribution function in $k$-space  for this model.

In conclusion, we have considered interaction effects in short QPCs
and shown that taking into account non momentum-conserving
processes, we can qualitatively account for the experimentally
observed behavior of the linear and non-linear conductance,
thermopower (including their magnetic field dependencies) and shot
noise at the so-called 0.7 anomaly. The gate voltage dependence can
also be explained within the present scheme, because the
backscattering is suppressed for larger values of $k_FL$. In the
high-temperature (but still $T\ll T_F$) non-perturbative regime, our
second-order self-consistent approach predicts that the conductance
approaches $\approx e^2/h$. It is an open and interesting problem to
verify this result by other non-perturbative methods.

We thank P. Brouwer, W. H\"ausler,  and J. Paaske for discussions.
This work was supported by the SFB TR 12 of the DFG and by the ESF
network INSTANS.


\begin{thebibliography}{99}

\bibitem{wharam}
D.A. Wharam {\sl et al.}, J. Phys. C {\bf 21}, L209 (1988);
B.J. van Wees {\sl et al.}, Phys. Rev. Lett. {\bf 60}, 848 (1988).

\bibitem{thomas}
K.J. Thomas, J.T. Nicholls, M.Y. Simmons, M. Pepper, D.R. Mace,
and D.A. Ritchie, Phys. Rev. Lett. {\bf 77}, 135 (1996).

\bibitem{kristensen} A. Kristensen {\sl et al.}, Phys. Rev. B {\bf 62}, 10950 (2000).

\bibitem{cronenwett}
S.M. Cronenwett {\sl et al.}, Phys. Rev. Lett.  {\bf 88}, 226805
(2002).

\bibitem{appleyard}
N.J. Appleyard, J.T. Nicholls, M. Pepper, W.R. Tribe, M.Y. Simmons,
and D.A. Ritchie, Phys. Rev. B {\bf 62}, R16275 (2000).

\bibitem{roche}
P. Roche {\sl et al.}, Phys. Rev. Lett. {\bf 93}, 116602 (2004);
L. DiCarlo {\sl et al.}, Phys Rev. Lett. {\bf 97}, 036810 (2006).

\bibitem{bruus}
H. Bruus, V.V. Cheianov, and K. Flensberg, Physica E {\bf 10}, 97
(2001); D.J. Reilly, Phys. Rev. B {\bf 72}, 033309  (2005).

\bibitem{richter}
A. Lassl, P. Schlagheck, and K. Richter, Phys. Rev. B {\bf 75},
045346 (2007).

\bibitem{meir}
Y. Meir, K. Hirose, and N.S. Wingreen, Phys. Rev. Lett. {\bf 89},
196802 (2002); T. Rejec and Y. Meir, Nature {\bf 442}, 900 (2006);
S. Ihnatsenka and I.V. Zozoulenko, cond-mat/0701657.

\bibitem{cornagliabalseiro}
P.S. Cornaglia and C.A. Balseiro, Europhys. Lett. \textbf{67}, 634
(2004); P.S. Cornaglia, C.A. Balseiro, and M. Avignon, Phys.
Rev. B {\bf 71}, 024432 (2005).

\bibitem{berggren}
C.K. Wang and K.-F. Berggren, Phys. Rev. B {\bf 54}, 14257(R)
(1996); A.A. Starikov, I.I. Yakimenko, and K.-F. Berggren, Phys.
Rev. B {\bf 67}, 235319 (2003).

\bibitem{seeligmatveev} G. Seelig and K.A. Matveev, Phys. Rev. Lett.
{\bf 90}, 176804 (2003)

\bibitem{matveev} K.A. Matveev, Phys. Rev. Lett. {\bf 92}, 106801 (2004);
M. Kindermann and P.W. Brouwer, Phys. Rev. B {\bf 74}, 125309 (2006).

\bibitem{meidanoreg}
D. Meidan and Y. Oreg, Phys. Rev. B \textbf{72}, 121312 (2005).

\bibitem{schmeltzer}
D. Schmeltzer, A. Saxena, A.R. Bishop, and D.L. Smith, Phys. Rev. B
\textbf{71}, 045429 (2005).

\bibitem{syljuaasen}
O.F. Sylju\aa sen, Phys. Rev. Lett. \textbf{98}, 166401  (2007).

\bibitem{sushkov}
C. Sloggett, A.I. Milstein, and O.P. Sushkov,  cond-mat/0606649.

\bibitem{glazman88}
L.I. Glazman and R.I. Shekhter, Zh. Eksp. Teor. Fiz. {\bf 94}, 292
(1988) [Sov.~Phys.  JETP {\bf 67}, 163 (1988)].

\bibitem{adia} Assuming a fully transmitting one-particle potential
is not in conflict with the condition $k_F L \sim 1$, since the
length $L$ is essentially given by the region with only one subband,
see also Ref.~\cite{sushkov}.

\bibitem{luttinger}
J.M. Luttinger, Phys. Rev. \textbf{121}, 942 (1961).

\bibitem{Binfty}
For a local interaction, the direct and exchange terms cancel to any
order in  $T/T_F$. However, for a non-local interaction the
cancellation tendency implies that scattering between equal spins is
to higher order in $(T/T_F)$ with a reduced prefactor due to the
cancellation.

\bibitem{thermodef}
H. van Houten, L.W. Molenkamp, C.W.J. Beenakker, and
C.T. Foxon, Semicond. Sci. Technol. {\bf 7}, B215 (1992).

\bibitem{lundeprl}
A.M. Lunde, K. Flensberg, and L.I. Glazman, Phys. Rev. Lett.
\textbf{97}, 256802 (2006).

\bibitem{footnew}
Again, in order to have perfect transmission at $T=0$, the
one-particle on-site energy must be canceled out, so that
$\epsilon_0=-\textrm{Re}\,[ \Sigma^r(0)]$ at zero temperature.
\end{thebibliography}
\end{document}